\documentclass[aps,prb,twocolumn,showpacs,amsmath,amssymb,floatfix]{revtex4}
\usepackage{epsfig}
\usepackage{bm}

\begin{document}

\title{Energy relaxation at quantum Hall edge}

\author{Ivan P. Levkivskyi and Eugene V. Sukhorukov}
\affiliation{D\'epartement de Physique Th\'eorique, Universit\'e de Gen\`eve, CH-1211 Gen\`eve 4, Switzerland}
\date{\today}

\begin{abstract}
In this work we address the recent experiments of Altimiras and collaborators, \cite{exper, exper2}
where an electron distribution function at the quantum Hall (QH) edge at filling factor $\nu=2$
has been measured with high precision. It has been reported that the energy of electrons injected 
into one of the two chiral edge channels with the help of a quantum point contact (QPC) is equally 
distributed between them, in agreement with earlier predictions, one being based on the Fermi gas 
approach, \cite{nigg} and the other utilizing the Luttinger liquid theory. \cite{giovan} 
We argue that the physics of the energy relaxation process at the QH edge may in fact be more 
rich, providing the possibility for discriminating between two physical pictures in experiment. 
Namely, using the recently proposed non-equilibrium bosonization technique \cite{our-phas} we 
evaluate the electron distribution function and find that the initial ``double-step'' distribution 
created at a QPC evolves through several intermediate asymptotics, before reaching eventual 
equilibrium state. At short distances the distribution function is found to be asymmetric 
due to non-Gaussian current noise effects. At larger distances, where noise becomes Gaussian, 
the distribution function acquires symmetric Lorentzian shape. Importantly, in the regime of low 
QPC transparencies $T$ the width of the Lorentzian scales linearly with $T$, in contrast to 
the case of equilibrium Fermi distribution, whose width scales as $\sqrt{T}$. Therefore, we propose 
to do measurements at low QPC transparencies. We suggest that the missing energy paradox \cite{exper2} 
may be explained by the non-linearities in the spectrum of edge states.

\end{abstract}

\pacs{73.23.-b, 03.65.Yz, 85.35.Ds}

\maketitle

\section{Introduction}

A two dimensional electron gas (2DEG) in strong perpendicular magnetic field exhibits the regime of quantum Hall 
effect\cite{qhe} (QHE). One of the peculiar phenomena specific to this regime is the appearance of  
one dimensional (1D)
{\em chiral} edge states, which are quantum analogs of skipping orbits. Recent extensive experimental 
studies\cite{firstMZ, Heiblum, Glattli, Litvin, Basel} 
of these sates have led to the emergence of a new field in condensed matter physics dubbed the electron optics.
On the theoretical side, there are two main points of 
view on the physics of quantum Hall (QH) edge states. One group of theories \cite{group1}  
suggests that at integer values of the Landau 
levels filling factor the edge excitations are free chiral \textit{fermions}. The second group of theories is 
based on the concept of the edge magneto-plasmon 
picture.\cite{group2} 
The fundamental edge excitations in these theories are the charged and neutral collective \textit{boson} modes.

The domain where these two approaches meet each other is the \textit{low-energy} effective theory.\cite{eff-theory} 
In the framework of this theory, both fermion and boson excitations are two forms of the same entity.
Namely, they can be equivalently rewritten in terms of each other:
$$
\psi(x,t) \sim \exp[i\phi(x,t)]
$$  
where $\psi(x,t)$ is the fermion field, and $\phi(x,t)$ is the boson field. 
However, this transformation is highly nonlinear, and in the presence of strong Coulomb interaction 
fermions are not stable and decay into the boson modes which are the eigenstates of the edge Hamiltonian.

Results of tunneling spectroscopy experiments \cite{Chang} reasonably agree with the free-electron
description of edge states. However, the first experiment on Aharonov-Bohm (AB)
oscillations of a current through the electronic Mach-Zehnder (MZ) interferometer \cite{firstMZ} has
shown that the phase coherence of edge states is strongly suppressed at energies, which are inversely 
proportional to the interferometer's size. Moreover, several subsequent experiments on MZ interferometers 
at filling factor $\nu = 2$ have shown puzzling results on finite bias dephasing\cite{Heiblum, Basel, Glattli, Litvin}
theoretically studied in \cite{Neder, Sim, Chalker,our,Sukh-Che}. 
Namely, the visibility of AB oscillations in these experiments is found to have a lobe-type pattern as a function of 
the applied voltage bias. Such results are difficult to explain in terms of the fermion picture, while they 
all follow naturally from the  plasmon physics,\cite{our} where the Coulomb interaction plays
a crucial role. Thus the boson picture of edge excitations might be more appropriate.

In contrast to mentioned above non-local experiments, some local measurements seem to be not able
to differentiate between two physical pictures of edge states. 
For example, both theories predict Ohmic behavior of the tunneling current, unless 
it is renormalized by a non-linear dispersion of plasmons.
Moreover, the equilibrium distribution of the bosons is equivalent to the one of  fermions 
(see the demonstration of this fact in Section \ref{sec-distr-long}). 
Therefore, it might be interesting to investigate non-equilibrium local properties of edge states.

\begin{figure}[h]
\epsfxsize=7cm
\epsfbox{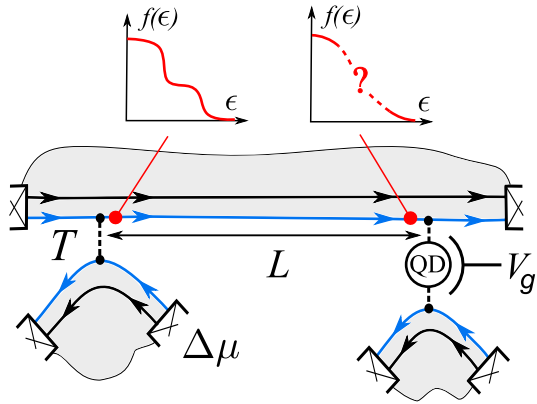}
\caption{Schematics of the experiments [\onlinecite{exper}] and [\onlinecite{exper2}]. 
The shaded region is filled by the 2D electron 
gas in the regime of the 
quantum Hall effect. At filling factor $\nu = 2$ there are two chiral edge states shown by the blue 
(the outer channel) and the black (the inner channel) lines. The QPC of the transparency $T$ and
 biased with the voltage difference $\Delta\mu$, injects electrons into the outer channel,
 and thus creates a non-equilibrium electron distribution. 
After the propagation along the QH edge, the distribution is detected at distances $L$ from the source
with the help of  a quantum dot with a 
single level controlled by the gate voltage $V_g$.   
} \vspace{-3mm}
\label{nu2}
\end{figure}

Non-equilibrium behavior of 1D systems has been a subject of intensive theoretical \cite{1d} 
and experimental \cite{1d-exp} studies for a long time. 
However, only recently it has become possible to measure an  electron distribution
 at quantum Hall edge $f(\epsilon)$  as a function of energy $\epsilon$ with high 
precision. \cite{exper} The main idea of the experimental  technique is to restore 
the function $f(\epsilon)$ by measuring the differential conductance $\mathcal{G}$ of tunneling 
between two 
edges through a single level in a quantum dot:
\begin{equation}
\mathcal{G}(\epsilon)\propto \partial f(\epsilon)/\partial\epsilon,
\end{equation}
where $\epsilon$ is the energy of the  quantum dot level, controlled by the gate voltage $V_g$. 
This technique has been used in experiments  [\onlinecite{exper2}] in order to investigate the 
energy relaxation at QH edge states 
at filling factor $\nu = 2$. 
The schematics of these experiments is shown in Fig.\ \ref{nu2}.
The main result is  that the electron distribution relaxes toward 
local equilibrium Fermi distribution, and the energy splits equally between the two edge channels.

The first theoretical models, based on the fermion picture [\onlinecite{nigg}] and on the plasmon approach 
[\onlinecite{giovan}], have come qualitatively to \textit{identical} conclusions. Namely, both works predict 
equal distribution of the energy between the edge channels, in agreement with the experimental findings.
In other words, based on the results of Refs.\ [\onlinecite{nigg}] and [\onlinecite{giovan}] alone,
the experimentalists are not able\cite{exper2} to discriminate between
two alternative descriptions of the physics of QH edge. Thus, it seems to be important and timely to
reanalyze the problem of the energy relaxation at the QH edge in order to make new, model specific and distinct
predictions that can be verified experimentally. This is exactly the purpose of the present work.

Here we show that the Coulomb interaction strongly affects the spectrum of collective edge excitations and leads to 
the formation of charged and dipole plasmons modes, which
propagate with different velocities.\cite{citeus} 
They carry away the energy of electrons injected through the QPC 
and equally
distribute it between edge channels at distances $L_{\rm ex}$ from the QPC. In addition 
to this observation, which agrees with findings of previous works, \cite{nigg,giovan} we stress that
the same process splits the wave packets of injected electrons, and leads to strong coupling of electrons
to the noise of the QPC current. The regime of weak injection, i.e., when the transparency 
of the QPC is small,  $T\ll 1$, deserves a special consideration. In this regime the current noise 
at relevant time scales becomes markovian, and as a result, the function $-\partial f(\epsilon)/\partial\epsilon$
acquires a Lorentzian shape. (This effect resembles a well known phenomenon of the homogeneous level broadening.)
Interestingly, the width of the Lorentzian scales as $T\Delta\mu$ at small $T$, where $\Delta\mu$ is the 
voltage bias applied to a QPC. In contrast, the width of the eventual equilibrium  Fermi distribution of 
thermalized electrons scales as $\sqrt{T}\Delta\mu$. If thermalization takes place at longer distances,
$L_{\rm eq}\gg L_{\rm ex}$, then the intermediate regime described here may be observed in experiment with
a weak injection. This would indicate that interactions strongly 
affects the physics at the edge and that the fermion picture becomes {\em inappropriate}.

In order to theoretically describe the experiments [\onlinecite{exper2}] and to quantitatively elaborate 
the  physical picture, we use the \textit{non-equilibrium} bosonization technique, which has been introduced 
in our previous work [\onlinecite{our-phas}]. The main idea of this approach is based on the fact that in a 1D chiral system one can find a non-equilibrium density matrix by solving equations of motion for plasmons with non-trivial boundary conditions. Then, one can rewrite an average over the non-equilibrium state of an interacting system 
in terms of the full counting statistics (FCS) generators\cite{Levitov} of the current at the boundary. In the situation considered in this paper, because of chirality of QH edge states,  interactions do not affect the 
transport through the QPC alone. This leads to a great simplification, because in the markovian limit the 
FCS generator for free electrons is known.\cite{Levitov} 

The structure of the paper is following: In Sec.\ \ref{sec-noneq} we describe the non-equilibrium 
bosonization technique in some details. Next, we use this technique in Sec.\ \ref{sec-corr} in order to find 
the electron correlation function for different distances from the QPC. Finally, we use these results
to find the electron distribution function in Sec.\ \ref{sec-distr}, and present our conclusions in 
Sec.\ \ref{sec-conclus}. Several important technical steps and the phenomena resulting from  the non-linearity 
of the spectrum of plasmons are described in  Sec.\ \ref{ap-energy} and Appendices.

\section{Non-equilibrium bosonization}
\label{sec-noneq}

We note, that the relevant energy scales in recent mesoscopic experiments with the QH edge state
\cite{exper,Heiblum,Basel,Glattli,Litvin} are very small. Therefore, it is appropriate to use the 
low-energy effective theory \cite{eff-theory} of the QH edge. One of the advantages of this theory 
is that it allows to take into account strong Coulomb interactions in a straightforward way.\cite{our}
However, an additional complication arises from the fact that in experiments \cite{exper,exper2} the 
injection into one of the two edge channels creates a strongly non-equilibrium state. We, therefore,
start by recalling in this section the method of non-equilibrium bosonization, proposed earlier in Ref.\ 
[\onlinecite{our-phas}], which is suitable for solving the type of a problem that we face. Throughout
the paper, we set $e=\hbar=1$.

\subsection{Fields and Hamiltonian}

According to the effective theory of QH edge,\cite{eff-theory} the collective fluctuations of the 
charge densities $\rho_{\alpha}(x)$ of the two edge channels, $\alpha= 1,2$, at filling factor $\nu=2$ 
are the only relevant degrees of freedom at low energies.
These charge densities may be expressed in terms of the \textit{chiral} boson fields, $\phi_{\alpha}(x)$, 
\begin{equation}
\rho_{\alpha}(x)=(1/2\pi)\partial_x\phi_{\alpha}(x),
\label{rho}
\end{equation}
which satisfy the following commutation relations:
\begin{equation}
[\phi_{\alpha}(x),\phi_{\beta}(y)]=i\pi\delta_{\alpha\beta}{\rm sgn}(x-y),
\label{fields}
\end{equation}
The vertex operator 
\begin{equation}
\psi_\alpha(x) = \frac{1}{\sqrt{a}}\,e^{i\phi_\alpha(x)}
\label{psi}
\end{equation}
annihilates an electron at point $x$ in the edge channel $\alpha$. The constant $a$ in the prefactor 
is the ultraviolet cutoff, which is not universal and will be omitted and replaced by other normalizations.
One can easily check, with the help of the commutation relations (\ref{fields}), that the operators
(\ref{psi}) indeed create a local charge of the value 1 at point $x$, and satisfy fermionic commutation
relations.

Close to the Fermi level the spectrum of electrons may be linearized, therefore the free-fermion 
part $\mathcal{H}_0$ of the total QH edge Hamiltonian $\mathcal{H} = \mathcal{H}_0 + \mathcal{H}_{\rm int}$
takes the following form:
\begin{equation}
\mathcal{H}_0 =-iv_F\sum\limits_{\alpha} \int dx \,\psi_\alpha^\dagger\partial_x\psi_\alpha,
\end{equation}
where the bare Fermi velocity $v_F$ is assumed to be the same for electrons at both edge channels. 
The second contribution to the edge Hamiltonian describes the density-density Coulomb interaction,
\begin{equation}
\mathcal{H}_{\rm int}  = (1/2)\sum_{\alpha,\beta}
\iint dx dy\, U_{\alpha\beta}(x-y) \rho_\alpha(x)\rho_\beta(y),
\label{Hamiltonian}
\end{equation}
which is assumed to be screened at distances $d$ smaller than the characteristic length scale $L$ 
in experiments [\onlinecite{exper,exper2,Heiblum,Basel,Glattli,Litvin}], i.e., $L\gg d$. Therefore, we may write:
\begin{equation}
U_{\alpha\beta}(x-y) = U_{\alpha\beta}\delta (x-y).
\label{short-r}
\end{equation}
Screening may occur due to the presence of either a back 
gate, or several top gates. We show below that the assumption (\ref{short-r}) results in the linear
spectrum of charge excitations. This approximation seems to be reasonable, agrees well with some
experimental observations such as an Ohmic behavior of the QPC conductance at low voltage bias,
and eventually does not strongly affect our main results. Nevertheless, below we relax this assumption
and investigate the effects of weak and strong non-linearities in the spectrum of charge excitations.  

After taking into account the relations 
(\ref{rho}) and (\ref{psi}) and applying the point splitting procedure, we arrive at the edge Hamiltonian 
of the quadratic form in boson fields:
\begin{equation}
\hspace*{-1pt}\mathcal{H} = \frac{1}{8\pi^2}\!\sum_{\alpha,\beta} V_{\alpha\beta}\!\!\int dx
\partial_x\phi_\alpha(x)\partial_x\phi_\beta(x),
\label{hamilt}
\end{equation}
which nevertheless contains free fermion contribution as well as the Coulomb interaction potential: 
\begin{equation}
V_{\alpha\beta} = 2\pi v_F \delta_{\alpha\beta} + U_{\alpha\beta}.
\label{kernel}
\end{equation}
Equations (\ref{fields}), (\ref{psi}), (\ref{hamilt}) and (\ref{kernel}) complete the description of 
the QH edge at low energies.

The experimentally found \cite{exper,exper2} electron distribution function at the outmost QH edge channel  
is given by the expression:
\begin{equation}
f(\epsilon) = \int dt e^{-i\epsilon t}\langle\psi_1^\dag(L,t)\psi_1(L,0)\rangle.
\label{distr-ferm}
\end{equation}
Rewriting this expression  via 
the boson fields we finally obtain 
\begin{subequations}
\begin{eqnarray}
f(\epsilon)& =&\int dt e^{-i\epsilon t}K(t),
\label{distr-phi-1}\\
K(t)& = &\langle e^{-i\phi_1(L,t)} e^{i\phi_1(L,0)}\rangle .
\label{distr-phi-2}
\end{eqnarray}\label{distr-phi}
\end{subequations}
where we have introduced the electron correlation function $K$, evaluated at
coincident points at distance $L$ from the QPC.
The proportionality coefficient in (\ref{distr-phi-2}) may be corrected later from the 
condition that $f(\epsilon)$ takes 
a value $1$ for energies well below the Fermi level (see, however, the discussion in Sec.\ \ref{ap-energy}
for further details).
In equilibrium, in order to evaluate the correlation function on the right hand side of this equation
one may now follow a standard procedure \cite{Giamarchi} of imposing periodic boundary conditions 
on the boson fields and diagonalizing the Hamiltonian (\ref{hamilt}).   
In our case, however, the average in (\ref{distr-phi}) has to be taken over a 
\textit{non-equilibrium} state created by a QPC. Attempting to express such a state entirely in 
terms of bosonic degrees of freedom is a complicated and not a best way to proceed.
We circumvent this difficulty by applying a non-equilibrium bosonization technique proposed 
in our earlier work [\onlinecite{our-phas}]. This technique is outlined below in some detail.

\subsection{Equations of motion, boundary conditions,\\ and FCS}
\label{NB}

The Hamiltonian (\ref{hamilt}), together with the commutation relations (\ref{fields}),
generates equations of motion for the fields $\phi_{\alpha}$, which have to be complemented with
boundary conditions:\cite{Ines} 
\begin{subequations}
\label{eom}
\begin{eqnarray}
\partial_t\phi_{\alpha}(x,t) = -\frac{1}{2\pi}\sum_\beta V_{\alpha\beta}\partial_x\phi_{\beta}(x,t),
\label{eoma}\\
\partial_t\phi_{\alpha}(0,t) = -2\pi j_{\alpha}(t).
\label{eomb}
\end{eqnarray}
\end{subequations}
The last equation follows from the charge continuity condition $\partial_t \rho_{\alpha}+\partial_x j_{\alpha}=0$ 
and the definition (\ref{rho}). Thus the operator $j_{\alpha}(t)$ describes a current through the 
boundary $x=0$ in the channel $\alpha$. For the convenience, we place a QPC in the outer channel $\alpha=1$ 
right before the boundary, so that the operator $j_{1}(t)$ describes an outgoing QPC's current.

The key idea of the non-equilibrium bosonization approach is to replace the average in Eq.\ (\ref{distr-phi}) 
by the average over temporal fluctuations of currents $j_{\alpha}$, the statistics of which is assumed
to be known. Indeed, although in general the fields $\phi_{\alpha}$ influence fluctuations of the currents $j_{\alpha}$, leading to such effects as the dynamical Coulomb blockade \cite{dynamicalCB} and 
cascade corrections to noise, \cite{nagaev} in the case of chiral fields describing QH edge states 
no back-action effects arise. \cite{Sukh-Che,our} As a consequence, at integer filling factors the electron 
transport through a single QPC is not affected by interactions, which seem to be an experimental fact. \cite{exper,Basel} Therefore, by solving equations  (\ref{eom}) one may express the correlation functions 
of the fields $\phi_{\alpha}$ in terms of the generator of full counting statistics (FCS):\cite{Levitov}
\begin{equation}
\chi_{\alpha}(\lambda,t)=\langle e^{i\lambda Q_{\alpha}(t)}e^{-i\lambda Q_{\alpha}(0)}\rangle.
\label{fcs}
\end{equation}
Here averaging is taken over \textit{free} electrons, and the operators 
\begin{equation}
Q_{\alpha}(t)=\int_{-\infty}^t dt' j_{\alpha}(t')
\label{Q}
\end{equation}
may be viewed as a total charge in the channel $\alpha$ to the right of the boundary at $x=0$.

To prove the connection of the electron correlations in (\ref{distr-phi}) to the generating
functions (\ref{fcs}), we come back to the discussion of the interaction effects, which are
in fact encoded in a solution of the equations of motion (\ref{eoma}). The long-range 
character of the Coulomb interaction leads to the logarithmic dispersion in the spectrum 
of collective charge excitations, the physical consequences of which are discussed in Secs.\
\ref{sec-distr} and \ref{ap-energy}. For a moment, to simplify equations (\ref{eoma}), we have assumed screening 
of the Coulomb potential at distances $d$ shorter than the characteristic length scale $v_F/\Delta\mu$, 
which is of the order of few microns in recent experiments. Nevertheless, it is very natural 
to assume that the screening length $d$ is much larger than the distance $a$ between edge channels, 
$d \gg a$, which does not exceed few hundreds nanometers. Therefore, one can write
\begin{equation}
U_{\alpha\beta} =\pi u,\quad u/v_F\sim\log(d/a)\gg 1,
\label{approx}
\end{equation}
i.e., the in-channel interaction strength is approximately equal to the intra-channel. 
As a result, the spectrum of collective charge excitations splits into two modes: a fast charged 
mode $\tilde\phi_1$ with the speed $u$, and a slow dipole mode $\tilde\phi_2$
with the speed $v \simeq v_F$. 

It is important to stress that the condition $d\gg a$, leading to (\ref{approx}),
results in a sort of universality: the solution of equations of motion  
(\ref{eoma}) in terms of the charged and dipole mode,
\begin{subequations}
\label{transform-phi}
\begin{eqnarray}
\phi_1(x,t) = \frac{1}{\sqrt{2}}[\tilde{\phi}_1(x-ut)+\tilde{\phi}_2(x-vt)]\\
\phi_2(x,t) = \frac{1}{\sqrt{2}}[\tilde{\phi}_1(x-ut)-\tilde{\phi}_2(x-vt)]
\end{eqnarray}
\end{subequations}
is only weakly sensitive to perturbations of our model, in particular to those that 
account for different bare Fermi velocities of edge channels and slightly different interaction strengths.

Applying now boundary conditions (\ref{eomb}) to the result (\ref{transform-phi}), we finally solve
equations of motion in terms of the boundary currents:
\begin{subequations}
\label{sol-full}
\begin{multline}
\phi_1(x,t) = -\pi\!\int_{-\infty}^{t_u}\!\!\!dt'[j_{1}(t')+j_{2}(t')]\\
-\pi\!\int_{-\infty}^{t_v}\!\!\!dt'[j_{1}(t')-j_{2}(t')],
\end{multline}\vspace{-25pt}
\begin{multline}
\phi_2(x,t) = -\pi\!\int_{-\infty}^{t_u}\!\!\!dt'[j_{1}(t')+j_{2}(t')]\\
+\pi\!\int_{-\infty}^{t_v}\!\!\!dt'[j_{1}(t')-j_{2}(t')],
\end{multline}
\end{subequations}
where we have introduced notations 
\begin{equation}
t_u = t- x/u,\quad t_v = t- x/v.
\label{notations}
\end{equation}
Finally, using the definition (\ref{Q}), we arrive at the solution 
in the compact form 
\begin{equation}
\phi_{1}(x,t)=-\pi[Q_{1}(t_u)+Q_{2}(t_u)+Q_{1}(t_v)-Q_{2}(t_v)],
\label{sol-fast}
\end{equation}
and to a similar expression for the inner channel. The physical meaning of this result is rather 
simple: when charges are injected into the channel $\alpha=1$ and $2$, they excite charged and 
dipole mode (note the minus sign in the fourth term on the right hand side) which have different 
propagation speeds $u$ and $v$. As a result, these charges arrive at the observation point $x$
with different time delays $x/u$ and $x/v$, and make a contribution to the field $\phi_1$ at 
different times (\ref{notations}).

\begin{figure}[h]
\epsfxsize=8cm
\epsfbox{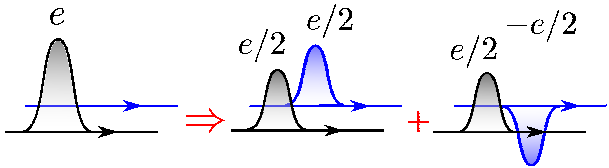}
\caption{Schematic illustration of the Coulomb interaction effect at the HQ edge at filling factor 
$\nu=2$. The electron wave-packet of the charge $e$ created in the outer edge channel (black line) 
decays into two eigenmodes of the Hamiltonian (\ref{hamilt}), the charged and dipole mode, which
propagate with different speeds and carry the charge $e/2$ in the outer channel. As a result, the 
wave packets do not overlap at distances larger than their width, and contribute independently to 
the electron correlation function with the coupling constant $\lambda = \pi$. \cite{footnote1} 
Similar situation arises when an electron is injected in the 
inner channel (blue line), however in this case the charged and dipole states carry opposite charges 
at the outer channel. Thus, there are four independent contributions to the correlation function 
in the outer edge channel. }
\vspace{-3mm} \label{fig-plasmon}
\end{figure}

When substituting this result into the correlation function in Eq.\ (\ref{distr-phi-2}) 
one may use the statistical independence of the current fluctuations at different 
channels and split the exponential functions accordingly: 
\begin{multline}
\label{corr-long}
K(t)=\langle e^{i\pi [Q_1(t_u)+Q_1(t_v)]}e^{-i\pi  [Q_1(t_u-t)+Q_1(t_v-t)]}\rangle 
\\ \times \langle e^{i\pi [Q_2(t_u)-Q_2(t_v)]}e^{-i\pi  [Q_2(t_u-t)-Q_2(t_v-t)]}\rangle .
\end{multline}
In the rest of the paper we will be interested in the correlation function at relatively long 
distances $L\gg v\tau_c$, where $\tau_c \simeq 1/\Delta\mu$ is the correlation 
time of fluctuations of the current through a QPC. (We show below that at this length
scale the energy exchange between two channels takes place.)
In this case, the partitioned charges $Q_\alpha$, taken at different times $t_u$ and $t_v$, 
are approximately not correlated, as illustrated in Fig.\ \ref{fig-plasmon}. 
This assumption is quite intuitive and may be easily  checked using Gaussian approximation.
We, finally, arrive at the following important result: 
\begin{equation}
\label{corr-short}
K(t) =\chi_1^2(\pi, t)\chi_2(-\pi,t)\chi_2(\pi,t),
\end{equation}
i.e., the electronic correlation function (\ref{corr-long}) may indeed be expressed in terms 
of the FCS generator (\ref{fcs}).

\section{Electron correlation function}
\label{sec-corr}

The expression (\ref{corr-short}) presents formally a full solution of the problem of evaluation of 
an electron correlation function. Generators of the FCS for free electrons in this expression, defined 
as (\ref{fcs}), may be represented as a determinant of a single particle operator,\cite{Levitov} 
and eventually, evaluated, e.g., numerically. However, a further analytical progress is possible
in a number of situations, which are important for understanding physics of the energy 
relaxation processes. In particular, we show in this section that for the case of equilibrium fluctuations  
of the boundary currents, the correlation function (\ref{corr-short}) as well as the electron
distribution function (\ref{distr-phi}) acquire an equilibrium free-fermionic form. The electron correlation 
function may also be found analytically away from equilibrium for the case of a Gaussian noise. Interestingly,
in the short-time limit, $t\ll 1/\Delta\mu$, the main contribution to the correlation function 
comes from zero-point fluctuations of boundary currents, and it behaves as a free-fermion correlator,
i.e., it scales as $1/t$. In the long-time limit, $t\gg 1/\Delta\mu$, the non-equilibrium
zero-frequency noise dominates, and the electron correlation function decays exponentially with time.
This is exactly the limit where a non-Gaussian markovian noise should also be taken into account. 

\subsection{Gaussian noise}
 
In the context of the noise detection physics \cite{Levitov,Edwards} the dimensionless counting variable $\lambda$ 
in the expression (\ref{corr-short}) for the FCS generator plays the role of a coupling constant. 
Typically, it is small, $\lambda\ll 1$, so that the contributions of high-order cumulants of current noise 
to the detector signal are negligible. \cite{Edwards} In contrast, in the physical situation that we consider 
in the present paper $\lambda=\pm\pi$, implying that the shape of the distribution function may be strongly
affected by high-order current cumulants. Nevertheless, it is instructive to first consider Gaussian fluctuations,
simply truncating the cumulant expansion at second order in $\lambda$. In this case the correlation function 
(\ref{corr-long}) may be evaluated exactly. The are many reasons for starting the analysis from considering
an example of a Gaussian noise: First of all, in equilibrium the current fluctuations in a 
chiral 1D system are always Gaussian. Second, as we show in the Appendix \ref{ap-supr}, the dispersion of the 
 charged and dipole modes leads to a suppression of higher order cumulants at large distances $L$. Finally, on 
 the Gaussian level it is more easy to investigate and compare contributions of zero-point fluctuations and 
 of non-equilibrium noise to the electron correlation function.  

Thus, expanding the logarithm of the right hand side of the Eq.\ (\ref{fcs})
to second order in $\lambda$ and accounting for the Eq.\ (\ref{Q}), we obtain
\begin{equation}
\log[\chi_{\alpha}(\lambda,t)]=i\lambda \langle j_{\alpha}\rangle t -\lambda^2J_{\alpha}(t).
\label{logchi}
\end{equation}
Here the Gaussian contribution of current fluctuations
$\delta j_\alpha(t) \equiv j_\alpha(t) - \langle j_\alpha\rangle$ is given by the following 
integral
\begin{equation}
J_{\alpha}(t) = \frac{1}{2\pi}\int\frac{d\omega S_{\alpha}(\omega)}{\omega^2+\eta^2}(1-e^{-i\omega t}),
\quad \eta\to 0,
\label{gaussian}
\end{equation}
where the non-symmetrized noise power spectrum is defined as
\begin{equation}
\label{S}
S_\alpha(\omega) = \int dt e^{i\omega t}\langle\delta j_\alpha(t)\delta j_\alpha(0)\rangle.
\end{equation}
In what follows, we apply this result for the evaluation of the electron correlation function 
in the case of equilibrium boundary conditions and in the case of a Gaussian noise far away  
from equilibrium.

\subsubsection{Equilibrium boundary conditions}
\label{b-f-eq}

One may propose the following simple test of the non-equilibrium bosonization method: Let us consider 
an infinite QH edge. In equilibrium, the charge densities and edge currents exhibit thermal fluctuations.
This is the case, in particular, for the currents $j_\alpha$ through the cross-section $x=0$, which are 
considered to be boundary conditions for the field $\phi_\alpha$ in our theory. Therefore, one may
evaluate the electron correlation function using these boundary conditions and compare it with the  
result of the standard equilibrium bosonization technique,\cite{Giamarchi} applied to a chiral 1D 
system.\cite{our} 

In equilibrium, $\langle j_{\alpha}\rangle=0$. The current noise power spectrum is given by the 
fluctuation-dissipation relation\cite{FDT}
\begin{equation}
S_\alpha(\omega)\equiv\int dt e^{i\omega t}\langle j_\alpha(t)j_\alpha(0)\rangle = \frac{1}{2\pi}\,
\frac{\omega}{1-e^{-\beta\omega}}.
\label{seq}
\end{equation}
Substituting this expression  into the equation (\ref{gaussian}), 
one obtains
\begin{equation}
\label{int-corr}
\log[\chi_\alpha(\lambda,t)] = -\frac{\lambda^2}{4\pi^2}
\int \frac{d\omega}{\omega}\frac{1-e^{-i\omega t}}{1-e^{-\beta\omega}}\,. 
\end{equation}
This integral may be evaluated by expanding the integrand in Boltzmann factors $e^{\pm\beta\omega}$
and integrating each term. Substituting the result (for $\lambda = \pi$) into Eq.\ (\ref{corr-short}), 
we arrive at the following expression for the electron correlation function 
in the case of equilibrium boundary conditions:
\begin{equation}
\label{corr-thermal}
K(t)\propto \frac{\beta^{-1}}{\sinh(\pi t /\beta)},
\end{equation}
which is, in fact, the equilibrium fermionic correlation function.
The straightforward calculations of the integral (\ref{distr-phi-1}) gives, naturally, 
the equilibrium distribution function, $f_1(\epsilon)=1/(1+e^{\beta\epsilon})\equiv  f_F(\epsilon)$, where 
we have fixed the normalization constant, as explained above. Thus for chiral, interacting quasi-1D systems
with a linear spectrum
equilibrium bosons also implies equilibrium distribution of fermions.

It is instructive to compare this result with the known expression for the electron correlation function
at $\nu=2$, found earlier in Ref.\ [\onlinecite{our}] with the help of the standard bosonization technique: 
\begin{equation}
\label{corr-known}
K(t)=\beta^{-1}\Big[\sinh\Big(\frac{x-y-vt}{v\beta/\pi}\Big)\sinh\Big(\frac{x-y-ut}{u\beta/\pi}\Big)\Big]^{-1/2}
\end{equation}
For $x = y$, details of the interaction leading to wave-packet splitting (see Fig.\ \ref{fig-plasmon}) 
vanish, and one obtains the expression (\ref{corr-thermal}), thus validating our approach.  
Moreover, the free-fermionic character of the correlation function at coincident points 
(\ref{corr-thermal}) justifies the assumption underlying the non-equilibrium 
bosonization procedure that the FCS generators (\ref{fcs}) may be taken as for free electrons.

\subsubsection{Gaussian noise away from equilibrium}

\begin{figure}[h]
\epsfxsize=8cm
\epsfbox{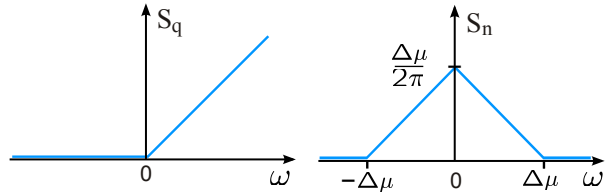}
\caption{Two contributions to the spectral density of noise (\ref{spectr}). 
{\em Left panel}: Quantum contribution $S_{\rm q}(\omega)$ generated by the 
incoming Fermi sea. This contribution vanishes at low frequencies $S_{\rm q}(0)= 0$, but dominates 
the behavior of the correlator (\ref{gaussian}) at short times, $t\Delta\mu\ll 1$. 
{\em Right panel}: Non-equilibrium contribution $S_{\rm n}(\omega)$ dominates at long times $t\Delta\mu\gg 1$,
i.e., in markovian limit.}
\vspace{-3mm} \label{noise}
\end{figure}

For a QPC far away from equilibrium, $\beta\Delta\mu\gg 1$, one may simply set the temperature to zero.
Straightforward calculations based on the scattering theory \cite{Buttiker} give the following 
result for the spectral density of noise (\ref{S}) of a QPC:
\begin{equation}
S_{\alpha}(\omega)= S_{\rm q}(\omega) + R_{\alpha} T_{\alpha} S_{\rm n}(\omega),
\label{spectr}
\end{equation}
where $T_\alpha=1-R_\alpha$ is the transparency of a QPC (i.e., the average occupation in 
the channel $\alpha$), $S_{\rm q}(\omega) =(1/2\pi)\omega  \theta(\omega)$ is the quantum, 
ground-state spectral function, and
$S_{\rm n}(\omega) =\sum_\pm S_{\rm q}(\omega\pm\Delta\mu)
-2S_{\rm q}(\omega),$ is the non-equilibrium contribution (see Fig.\  \ref{noise}). 
Note, that the noise power 
(\ref{spectr}) differs from the one for a non-chiral case. \cite{Edwards}

Evaluating the integral (\ref{gaussian}), we arrive at the following expression
\begin{equation}
J_{\alpha}(t) = (1/4\pi^2)[\log t + 2R_\alpha T_\alpha f(\Delta\mu t)],
\end{equation}
where the logarithm of time originates from the quantum contribution $S_{\rm q}$, and the 
dimensionless function $f(\Delta\mu t)$, describing non-equilibrium noise, is given by
the integral
\begin{equation}
\label{f}
f(\Delta\mu t) = \int_0^{1}\!\!\!ds\,\frac{1 - s}{s^2}\,[1-\cos(\Delta\mu s)].
\end{equation}
This function has a quadratic behavior $f(\Delta\mu t) \simeq (\Delta\mu t)^2/4$ at short times $\Delta\mu t\ll 1$, 
while in the long-time (markovian) limit, $\Delta\mu t\gg 1$, the dominant contribution to this function is 
linear in time: $f(\Delta\mu t)\simeq (\pi/2)|\Delta\mu t|$.

For the purpose of further analysis we need the electron correlation function in the long-time limit.
Taking into account that $\langle j_\alpha\rangle = \Delta\mu T_\alpha /2\pi$, we find the cumulant 
generating function
\begin{eqnarray}
\log [\chi_\alpha
] &=& \frac{i\lambda}{2\pi} \Delta\mu T_\alpha t \nonumber
\\ &-&\Big(\frac{\lambda}{2\pi}\Big)^2(\log t - \pi R_\alpha T_\alpha |\Delta\mu t|),\;\, \Delta\mu t\gg 1.\quad
\end{eqnarray}
Finally, substituting this result into the equation (\ref{corr-short}), and setting $T_1=T$ and 
$T_2=1$ according to the situation shown in Fig.\ \ref{nu2}, we obtain the electron correlation function
in the long-time limit:
\begin{equation}
K(t) \propto t^{-1}e^{i\Delta\mu T t- \pi RT\Delta\mu|t|/2},
\quad\Delta\mu t\gg 1.
\label{cf-gauss}
\end{equation}
Note, that the expression (\ref{cf-gauss}) contains the quantum contribution in the form of a single pole,
as for free fermions, as well as the non-equilibrium contribution in the form of an exponential envelop, 
those width is determined by the noise power at zero frequency, $S_1(0) = RT\Delta\mu/2\pi$. 
The phase shift of the correlator is determined by the ``average'' voltage bias 
$\langle\Delta\mu\rangle=\Delta\mu T$
of the incoming stream of electrons, diluted by the QPC. In the next section we show that this
mean-field like effect of the dilution is strongly modified by a non-Gaussian component of noise.

\subsection{Non-Gaussian markovian noise}

Here we consider non-Gaussian noise and show that the contribution of high-order cumulants of current
to the correlation function is not small. Note, that the quantum ground state part of the current noise, 
$S_{\rm q}$, that dominates at short times, is pure Gaussian. Therefore, the denominator in the expression
(\ref{cf-gauss}) remains unchanged. In the long time, markovian limit, the dominant
contribution to the FCS generator comes from the non-equilibrium part of noise,
which, e.g., is described by $S_{\rm n}$ in Gaussian case. For a QPC, the markovian FCS generator is given 
by the well known expression \cite{Levitov} for a Binomial process:
\begin{equation}
\chi_{1}(\lambda,t)=(R+Te^{i\lambda})^N,
\end{equation}
where $N=\Delta\mu t/2\pi$ is the total number of electrons that contribute to noise.
Applying the analytical continuation $\lambda\to \pi$, we obtain
\begin{equation}
\log[\chi_{1}(\pi,t)]=\frac{\Delta\mu t}{2\pi}\big[\log|T-R|+i\pi\theta(T-R)\big].
\label{logchi-qpc}
\end{equation}
Substituting this expression to the correlation function (\ref{corr-short}), we arrive 
at the result
\begin{equation}
\label{cf-nongauss}
K(t)\propto t^{-1}e^{i\theta(T-R)\Delta\mu t +\log|T-R|\Delta\mu|t|/\pi},
\end{equation}
where the imaginary part of the exponent determines the effective voltage bias, while the 
real part is responsible for dephasing.

Interestingly, at the point $T = 1/2$  the dephasing rate is divergent, and the effective 
voltage bias drops to zero abruptly for $T<1/2$. It has been predicted in Ref.\ [\onlinecite{our-phas}]
that this behavior may lead to a phase transition in the visibility of Aharonov-Bohm oscillations
in electronic Mach-Zehnder interferometers. We will argue below that no sharp transition 
arises in the electron distribution function. However, it leads to its strong asymmetry with 
respect to the average voltage bias $\langle\Delta\mu\rangle=T\Delta\mu$ of the outer channel.

\section{Electron distribution function}
\label{sec-distr}

In this section we use the results (\ref{corr-thermal}), (\ref{cf-gauss}) and (\ref{cf-nongauss}) 
for the correlation function of electrons to evaluate and analyze the electronic distribution function. 
We start by noting that the experiments [\onlinecite{exper,exper2}] are done in a particular regime
of strong partitioning $T\approx 0.5$ at the QPC injecting current to the channel $\alpha=1$. This
detail, which seem to be irrelevant from the first glance, is in fact of crucial importance. 
Indeed, as it follows from the expressions (\ref{cf-gauss}) and (\ref{cf-nongauss}), the main contribution 
to the integral (\ref{distr-phi-1}) for the correlation function comes from times $t$ of the order of 
the correlation time $\tau_c\simeq 1/\Delta\mu$, 
where our results based on the markovian noise approximation are, strictly speaking,  not valid. However, the numerical calculations
show that the non-equilibrium distribution in this regime is very 
close to the equilibrium one. Therefore, the actual equilibration of electrons, which is reported in the experiment 
[\onlinecite{exper}] to occur at distances $v/\Delta\mu$, may in fact take place at even longer distances 
$L_{\rm eq}\gg v/\Delta\mu$ due to an unknown mechanism (not considered here).

Indeed, if the chiral LL model considered in our paper is valid, then neither the strong interaction between 
electrons of two edges taken alone, nor the weak dispersion of plasmons resulting from a long-range character 
of Coulomb interaction may lead to the equilibration, because the systems remain integrable. Thus it seems to 
be reasonable to assume that the equilibration length $L_{\rm eq}$ may indeed be quite long. Therefore, in order 
to explore the physics of collective charge excitations at intermediate distances we propose to consider 
a regime of weak injection at the QPC: $T\ll 1$. Firstly, we 
note that in this case our results (\ref{cf-gauss}) and (\ref{cf-nongauss}) may indeed be used to evaluate 
the electron distribution function, because the main contribution to the integral (\ref{distr-phi-1})
arises from markovian time scales. Secondly, and more importantly, in this regime the electron distribution function 
acquires a strongly non-equilibrium form and the width of the order of $T\Delta\mu$, which plays a role of the new 
energy scale. Moreover, the advantage of the weak injection regime is that it allows to investigate a non-trivial 
evolution of the distribution function, which arises due to bosonic, collective character of excitations and    
goes via several well distinguishable steps.

\subsection{Short distances}

At distances of the order of the energy exchange length scale 
\begin{equation}
\label{len-nongauss}
L_{\rm ex}\equiv v/\Delta\mu 
\end{equation}
the initial double step distribution function is strongly perturbed by the interaction between channels. 
As we argued in Sec.\ \ref{NB}, at distances $L\gg L_{\rm ex}$ the charged and dipole modes split and 
make independent contributions to the electron correlation function. Therefore, we may rely on the result 
(\ref{cf-nongauss}). Applying the limit $T\ll 1$ to this expression  and evaluating the Fourier transform, 
we find:
\begin{equation}
\label{distr-nongauss}
-\frac{\partial f(\epsilon)}{\partial\epsilon} = \frac{\Gamma_{\rm ng}/\pi}{\epsilon^2 + \Gamma_{\rm ng}^2},
\quad \Gamma_{\rm ng}=2T\Delta\mu/\pi.
\end{equation}
Here, the missing prefactor in the correlation function has been fixed by the requirement that $f(\epsilon)=1$
at $\epsilon\to-\infty$. Thus, we conclude that energy derivative of the distribution function 
acquires a narrow Lorentzian peak, which is shifted with respect to the average bias $\langle\Delta\mu\rangle=T\Delta\mu$ 
and centered at $\epsilon=0$. The last effect is a unique signature of the non-Gaussian character of noise.
Because of the electron-hole symmetry of the Binomial process, in the limit $R\ll 1$ the Lorentzian peak 
obviously has a width $\Gamma_{\rm ng}=2R\Delta\mu/\pi$ and centered at $\epsilon=\Delta\mu$.

We stress again, that the result (\ref{distr-nongauss}) holds only for small enough energies 
close to the Fermi level, namely, for $|\epsilon| < \Delta\mu$, where the main contribution arises from the 
noise in markovian limit. In fact, the result (\ref{distr-nongauss}) fails at large energies in somewhat non-trivial 
way. Namely, it is easy to see that any electron distribution function has to satisfy
the sum rule 
\begin{equation}
\langle\Delta\mu\rangle\equiv\epsilon_0+\int\limits_{\epsilon_0}^\infty d\epsilon\,f(\epsilon)
=-\int\limits_{-\infty}^\infty d\epsilon\, \epsilon\, \partial f(\epsilon)/\partial\epsilon,
\label{sum-rule1}
\end{equation}
where $\epsilon_0$ is the cutoff well below the Fermi level, 
and the ``average'' bias $\langle\Delta\mu\rangle=T\Delta\mu$ in the case of linear dispersion of plasmons.
This sum rule simply expresses the requirement of the conservation of the charge current 
and implies certain amount of asymmetry in the 
distribution function.
In the present case, such an asymmetry arises in the power-law tails of the function $-\partial f(\epsilon)/\partial\epsilon$ 
and originates from quantum non-equilibrium noise. It can be seen in Fig.\ \ref{fig-distrib}, where the results of numerical 
calculations are shown. 

Moreover, at energies of the order of $\Delta\mu$ the power-law behavior 
of the function (\ref{distr-nongauss}) has to have a cut-off, because the QPC does not provide 
energy much larger than 
the voltage bias. Quantitatively, this follows from the conservation of the
energy. We demonstrate below that for the system with linear dispersion of plasmons, the heat flux in 
the outer channel can be written entirely in terms of the single-electron distribution
function (in unites $e=\hbar=1$), 
\begin{equation}
I_{\rm m} = (1/2\pi)\int d\epsilon\,\epsilon\,[f(\epsilon)-\theta(\langle\Delta\mu\rangle-\epsilon)],
\label{flux-def}
\end{equation}
as in the case of free electrons. We use the subscript ``${\rm m}$'' in order to emphasize
the fact that it is this quantity that has been measured in the experiment [\onlinecite{exper2}]. 
In Sec.\ \ref{ap-energy} we show that 
at distances $L\gg L_{\rm ex}$ the total heat flux injected at a QPC splits equally between two edge channels,
therefore integrating Eq.\ (\ref{flux-def}) by parts and substituting the heat flux for a double-step distribution, 
we obtain 
\begin{equation}
I_{\rm m} = -\frac{(T\Delta\mu)^2}{4\pi}-\frac{1}{4\pi}\int d\epsilon\, \epsilon^2 \frac{\partial f(\epsilon)}{\partial\epsilon}=\frac{TR(\Delta\mu)^2}{8\pi}
\label{sum-rule2}
\end{equation}
for $L\gg L_{\rm ex}$.
One can see from Eq.\ (\ref{distr-nongauss}) that indeed the power-law behavior has to have a cut-off at $|\epsilon|\sim\Delta\mu$.
We stress, however, that this summation rule is less universal than the one given by Eq.\ (\ref{sum-rule1}), because it accounts 
only a single-particle energy of electrons and fails in the case of a non-linear spectrum of plasmons, considered in Sec. \ref{ap-energy}
in detail.

\subsection{Intermediate distances}
\label{sec-distr-interm}

So far we have considered the case of a linear spectrum of plasmons. This is a reasonable assumption, taking into
account the fact that non-linear corrections in the spectrum of plasmons lead to a nonlinear corrections
in the Ohmic conductance of a QPC. However, the experiments [\onlinecite{exper,exper2}] seem to be done 
in the Ohmic regime. Nevertheless, even in the case of a weak non-linearity in the spectrum of the the both modes 
of the sort\cite{footnote2} 
\begin{equation}
k_j(\omega) = \omega/v_j + \gamma_j \omega^2{\rm sign}(\omega),\quad v_1=u,\; v_2=v,
\label{spectrum}
\end{equation}
barely seen in the conductance of a QPC, an intermediate length scale $L_{\rm g}$ may arise at which high-order cumulants
of current are suppressed, and the noise becomes effectively Gaussian. This situation occurs 
when the wave packets of the original width $v/(T\Delta\mu)$ overlap. A simple estimate using the nonlinear correction 
(\ref{spectrum}) gives the length scale 
\begin{equation}
L_{\rm g}=1/\gamma(T\Delta\mu)^2, \quad \gamma\equiv \min(\gamma_j).
\label{len-gauss}
\end{equation}
We support this conclusion by rigorous calculations in Appendix \ref{ap-supr}.

\begin{figure}[h]
\epsfxsize=8cm
\epsfbox{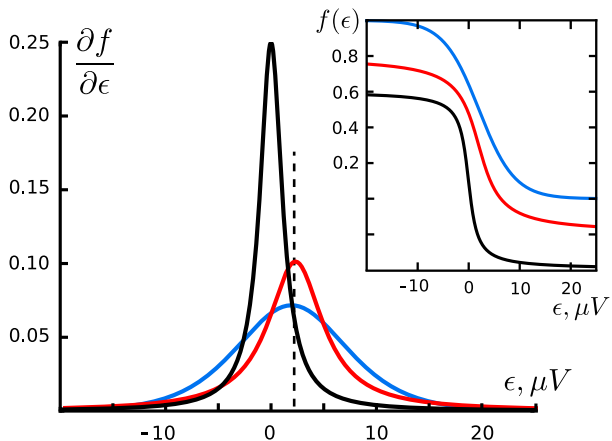}
\caption{Energy derivative of the electron distribution function, $-\partial f/\partial\epsilon$, 
is shown for different distances $L$ from the QPC injecting current. The transparency of the QPC
is set to $T = 0.05$ and voltage bias is $\Delta\mu = 40\, \mu V$. {\em Black line}:
$-\partial f(\epsilon)/\partial\epsilon$ for short distances $L_{\rm ex}\ll L\ll L_{\rm g}$, 
so that the noise is Gaussian (\ref{distr-nongauss}).  {\em Red line}: 
$-\partial f(\epsilon)/\partial\epsilon$ for intermediate distances $L_{\rm g}\ll L\ll L_{\rm eq}$, 
where the noise is Gaussian (\ref{distr-gauss}).  
{\em Blue line}: The derivative of the  Fermi distribution function at the temperature given by Eq.\ (\ref{fermi-width}). 
The dashed line is a guide for eyes
at the energy equal to the effective voltage bias $\langle\Delta\mu\rangle = T\Delta\mu = 2\, \mu V$. 
{\em Inset}: The same distribution functions are shown in the integrated form.
They are shifted vertically by $0.2$ intervals for clarity.}
\vspace{-3mm} \label{fig-distrib}
\end{figure}

The non-linearity in the spectrum is weak, 
if $\gamma vT\Delta\mu\ll 1$. This implies that $L_{\rm g}\gg L_{\rm ex}$, and leads 
to the possibility to observe non-Gaussian effects at distances $L_{\rm ex}\ll L\ll L_{\rm g}$,
discussed in the previous section. Obviously, the same requirement
also guarantees that dispersion corrections to the Ohmic conductance of a QPC are small. This allows us to 
neglect corrections to the quantum part of the electron correlation function and to use the result 
(\ref{cf-gauss}) for a Gaussian noise. Substituting this result
to the equation (\ref{distr-phi-1}), we obtain
\begin{equation}
\label{distr-gauss}
-\frac{\partial f(\epsilon)}{\partial\epsilon} 
= \frac{\Gamma_{\rm ng}/\pi}{(\epsilon-\langle\Delta\mu\rangle)^2 + \Gamma_{\rm g}^2},
\quad \Gamma_{\rm g}=\pi T\Delta\mu/2,
\end{equation}
in the case $L_{\rm g}\ll L\ll L_{\rm eq}$. 
One can see, that the width of the function (\ref{distr-gauss}) 
is almost twice as large compared to the one the function (\ref{distr-nongauss}). 
Moreover, the function (\ref{distr-gauss}) satisfies the sum rule (\ref{sum-rule1}). Therefore, we do not 
expect any asymmetry in the high-energy tails of this function, in contrast to the situation with the 
non-Gaussian noise. The comparison of distribution functions in these two regimes is shown in 
Fig.\ \ref{fig-distrib}. 

So far we have considered a situation where both charged and dipole mode are dispersive. 
If for some reason the dispersion of one of the modes is negligible,   
then higher order cumulants are suppressed only by a factor of two. 
The derivative of the electron distribution function in this situation is given
by the Lorentzian 
\begin{equation}
 \frac{\partial f(\epsilon)}{\partial\epsilon} = \frac{(\Gamma_{\rm ng}+\Gamma_{\rm g})/ 2\pi}{(\epsilon 
- \langle\Delta\mu\rangle/2)^2 + (\Gamma_{\rm ng}+\Gamma_{\rm g})^2/4}
\end{equation}
centered at $\langle\Delta\mu\rangle/2 = \Delta\mu T /2$ with the width
$(\Gamma_{\rm ng}+\Gamma_{\rm g})/2 = (1/\pi +\pi/4)\Delta\mu T$. This is because 
one mode brings only Gaussian component of the markovian noise, while the other one
brings full non-Gaussian noise.

\subsection{Long distances}
\label{sec-distr-long}

Next, we consider the distribution function at long distances, $L\gg L_{\rm eq}$,
after the equilibration takes place. The temperature of the eventual equilibrium distribution
may be found from the conservation of energy. In the next section we show that the heat flux produced
at QPC splits equally between two edge states. In the situation of linear dispersion 
the distribution function acquires the form
\begin{equation}
f(\epsilon)=\frac{1}{1+e^{(\epsilon-\langle\Delta\mu\rangle)/\Gamma_{\rm eq}}}.
\label{eq-distr}
\end{equation}
The possibility of such equilibration process is suggested by the fact
that the equilibrium distribution of bosons implies the equilibrium distribution
for electrons, as has been shown in Sec.\ \ref{b-f-eq}. Obviously, the distribution (\ref{eq-distr})
satisfies the sum rule (\ref{sum-rule1}), while the energy conservation condition (\ref{sum-rule2})
may  now be used in order to find the effective temperature:
\begin{equation}
\label{fermi-width}
\Gamma_{\rm eq} = \sqrt{3T/2\pi^2}\Delta\mu,
\end{equation}
where we have used $T\ll 1$.

We conclude that the  width of the equilibrium distribution scales as $\sqrt{T}$, in contrast 
to the case of a non-equilibrium distribution at shorter distances from the current source, 
where it scales linear in $T$. Therefore, if $T$ is small, an equilibrium and non-equilibrium 
distributions may easily be distinguished, as illustrated in Fig.\ \ref{fig-distrib}. 
In the situation where the dispersion can not be neglected, the equilibrium distribution of fermions
is not given by the Fermi function (\ref{eq-distr}). This situation 
deserves a separate consideration, which is provided in the next section.

\section{Measured and total heat fluxes}
\label{ap-energy}

We have seen that in the case of weakly dispersive plasmons, $\gamma v T\Delta\mu\ll 1$,
the non-linearity in the spectrum leads to the suppression of high-order cumulants of current
noise at relatively long distances, which strongly affects the distribution function. On
the other hand, the direct contribution of the non-linear correction in the spectrum to 
local physical quantities, such as the QPC conductance and the heat flux, is small and has been
so far neglected.  Nevertheless, it may manifest itself experimentally in a quite remarkable
way. In this section we show that the non-linearity in the the plasmon spectrum contributes
differently to the measured heat flux (\ref{flux-def}) and to the actual heat flux expected from the 
simple evaluation of the Joule heat. As we demonstrate below, this may, under certain circumstances, 
explain the missing
energy paradox in the experiment [\onlinecite{exper2}].  

We start by noting that the measured flux (\ref{flux-def}) at the distance $L$ form the QPC
may be expressed entirely in terms of the excess
noise spectrum ${\mathbb S}_\alpha(\omega)\equiv S_\alpha(\omega)-(1/2\pi)\omega\theta(\omega)$
of edge channels right after the QPC, where $S_\alpha(\omega)$ is defined in (\ref{S}). Namely,
in Appendix \ref{ap-energy-ap} we derive the following result:
\begin{multline}
\label{tot-energy}
I_{\rm m}(L) =\frac{1}{4}\int_{-\infty}^\infty d\omega 
\{{\mathbb S}_1(\omega)[1+\cos(\Delta k L)] \\ + {\mathbb S}_2(\omega)[1-\cos(\Delta k L)]\},
\end{multline}
where $\Delta k\equiv k_1(\omega)-k_2(\omega)$, and  $k_j(-\omega)=-k_j(\omega)$. 
Importantly, this result holds for an arbitrary non-linear spectrum $k_j(\omega)$ of the charged and dipole modes,
and for a  non-Gaussian noise in general, i.e., high-order cumulants simply do not contribute. 

One can easily find two important limits of Eq.\ (\ref{tot-energy}): 
for $L = 0$ we immediately obtain an expected result 
\begin{equation}
I_{\rm m}(0) = \frac{1}{2}\int_{-\infty}^{\infty} d\omega \,{\mathbb S}_1(\omega), 
\end{equation}
while at $L\to\infty$ the cosine in (\ref{tot-energy}) acquires
fast oscillations, and we get 
\begin{equation}
I_{\rm m}(\infty) = \frac{1}{2}\int_{-\infty}^{\infty} d\omega [{\mathbb S}_1(\omega)+{\mathbb S}_2(\omega)].
\label{I-at-infty}
\end{equation}
To be more precise, this happens at $L\gg L_{\rm ex}=v/\Delta\mu$.
At zero temperature ${\mathbb S}_2$ vanishes, and 
the single-electron heat flux $I_{\rm m}$, created at the QPC, splits equally between 
edge channels: $I_{\rm m}(\infty)=I_{\rm m}(0)/2$. Note also, that in the case of 
linear dispersion ${\mathbb S}_\alpha=T_\alpha R_\alpha S_{\rm n}$, where $S_{\rm n}$ 
is shown in Fig.\ \ref{noise}.

In the next step, we rewrite the same measured flux in terms of the plasmon distributions
$n_j(k)=\langle \tilde a^\dagger_j(k)\tilde a_j(k)\rangle$,
see Appendix \ref{ap-energy-ap}:
\begin{eqnarray}
I_{\rm m}(\infty) &=& \frac{1}{4\pi}\sum_j\int\limits \frac{dk}{k} 
\, \omega_j^2(k)n_j(k) + I_{\rm q},
\label{measured-f}\\
I_{\rm q} &=& \frac{1}{8\pi}\sum_j\int \frac{dk}{k} 
[\omega_j^2(k)-(v_jk)^2],
\label{quantum-f}
\end{eqnarray}
where $v_j=\partial\omega_j/\partial k$ are the plasmon speeds at $k=0$. 
The term $I_{\rm q}$ 
is the contribution to the measured flux
from quantum smearing of the zero-temperature electron distribution function $f_0(\epsilon)$ 
close to the Fermi level, which originates from a non-linear dispersion of plasmons.

Here an important remark is in order.  
The integral (\ref{quantum-f}) may diverge at large $k$ and has to be cut off at the upper
limit. This is because there is no guarantee of the free-fermionic behavior of the correlator $K(t)$
at short times and of the zero-temperature 
electron distribution function $f_0(\epsilon)$ at large energies. Thus, the integral (\ref{flux-def}) has 
to be also cut off, which is what in fact is done in experiment. 
In contrast, the spectrum of plasmons is linear at small $k$, and thus the distribution 
function $f_0(\epsilon)$ has a free-fermionic behavior close to the Fermi level.   
Our definition of  $I_{\rm q}$ corresponds to the normalization of $f_0(\epsilon)$ to have a discontinuity of the value $-1$ 
at $\epsilon=\langle\Delta\mu\rangle$.
The experimentally measured
$I_{\rm q}$ may differ from the one defined in (\ref{quantum-f}) by a constant,
which is, on the other hand, independent on the voltage bias $\Delta\mu$.

Next, we note that the actual total heat flux in the case of a non-linear dispersion of plasmons
acquires the completely different form \cite{footnote3}
\begin{equation}
I_{\rm h}= \frac{1}{2\pi}\sum_j\int dk \frac{\partial\omega_j}{\partial k}\omega_j(k)\,n_j(k),
\label{actual-f}
\end{equation} 
and thus in general $I_{\rm m}\neq I_{\rm h}/2$, contrary to what has been assumed 
in the experiment [\onlinecite{exper2}].
This may explain the missing energy paradox. Indeed, assuming the low $\omega$ spectrum 
of the general form 
\begin{equation}
k_j = \omega/v_j + \gamma_j \omega^{\ell_j},\quad j=1,2,
\end{equation}
where $\gamma_j$ are small,
and equilibration of plasmons at $L\to \infty$, i.e., 
$n_j(k)=n_B(\omega_j/\Gamma_{\rm eq})=1/[\exp(\omega_j/\Gamma_{\rm eq})-1]$, 
we obtain the missing heat flux as
\begin{equation}
I_{\rm m}-I_{\rm q}-I_{\rm h}/2=\sum_{j=1,2}c_j\gamma_jv_j\Gamma_{\rm eq}^{\ell_j+1},
\label{paradox}
\end{equation}
where the constants $c_j=(1/4\pi)\int dz z^{\ell_j+1}n_B(z)$ are of the order of $1$.
This result implies that experimentally, the missing heat flux may be found investigating 
its bias dependence and the spectrum of plasmons.

\begin{figure}[h]\begin{center}
\epsfxsize=5cm
\vspace{3mm}
\epsfbox{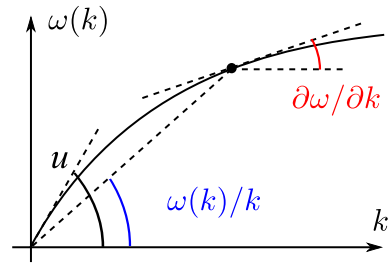}
\caption{Typical spectrum of charged plasmon in the case of the 
Coulomb interaction screened at distances $d\ll 1/k$. 
This spectrum is concave, i.e.,  
$\partial\omega/\partial k < \omega(k)/k$.}
\vspace{-4mm} \label{fig-coulomb}\end{center}
\end{figure}

Let us consider an example where  the dispersion of charged plasmon at small $k$ arises 
from the screened Coulomb interaction: \cite{group2}
\begin{multline}
\label{coulomb-exact}
\omega/k = 2[\mathcal{K}_0(ka)-\mathcal{K}_0(kd)]\\
=2\ln(d/a)-(1/2)(kd)^2\ln(2/kd),
\end{multline}
where $a$ is the high-energy cutoff, $d$ is the distance to the gate, such as $ka\ll kd\ll 1$,
and $\mathcal{K}_0$ is the MacDonald function. The low-$k$ asymptotics of this spectrum is 
illustrated in Fig.\ \ref{fig-coulomb}. One can see that the spectrum is concave, 
so that in this case the measured heat flux (\ref{measured-f}) is larger than the half of the actual 
heat flux (\ref{actual-f}). In addition, the effect is weak, because $kd\sim 0.1$ in the experiment 
[\onlinecite{exper2}]. Thus, the dispersion of the Coulomb interaction potential alone is not 
able to explain the missing flux paradox. Various mechanisms of convex dispersion are still possible and will be 
investigated elsewhere.

\section{Conclusions}
\label{sec-conclus}

Earlier theoretical works on quantum Hall edge states at integer filling factors 
may be divided into two groups: fermion based and boson 
based theories. Recent interference experiments suggest that the boson approach might be more appropriate for 
the description of the edge physics. However, both groups of theories give the same predictions for the local 
physical quantities at equilibrium. Moreover, the first theoretical works based on fermion \cite{nigg}
and boson \cite{giovan} approaches and addressing the 
non-equilibrium local measurements, have not been able to make qualitatively distinct predictions.  
In this paper we show that it is nevertheless \textit{possible} to test and differentiate between  
two approaches with the local non-equilibrium measurements.

We address recent experiments [\onlinecite{exper,exper2}] with quantum Hall edge states at filling factor 2, 
where an energy relaxation process has been investigated by creating a non-equilibrium state 
at the edge with the help of a QPC and reading out the electron distribution downstream using a 
quantum dot. We use the non-equilibrium bosonization approach \cite{our-phas} in order to describe the gradual relaxation 
of initially non-equilibrium, double-step electron distribution to its equilibrium form.
In the framework of this approach the non-equilibrium initial state is encoded 
in the boundary conditions for the equations of motions that depend on interactions. 
We show that the electrons excite two plasmons: fast charged and slow dipole mode. Thus the electron 
correlation function (\ref{corr-short}) is expressed in terms of the four contributions, each having form 
of the generator of FCS of free electrons with the coupling constant $\lambda = \pi$. Evaluating the 
Fourier transform of this function, we find the electron distribution function.

\begin{figure}[h]\begin{center}
\epsfxsize=9cm
\epsfbox{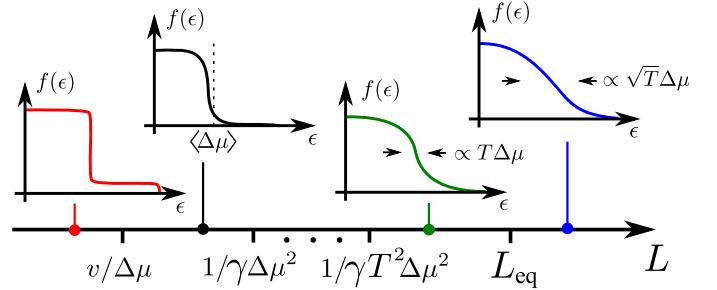}
\caption{Different length scales for energy relaxation processes and corresponding distribution functions 
in each regime are schematically shown. {\em Red curve}: The initial double-step distribution function. 
{\em Black curve}: At distances $L\gg \Delta\mu/v$ the distribution function is strongly asymmetric with respect 
to the ``average'' bias $\langle\Delta\mu\rangle = T\Delta\mu$. {\em Green curve}: 
At distances $L\gg 1/\gamma (T\Delta\mu)^2$ the distribution function is a Lorentzian with the width 
that scales as $T\Delta\mu$. {\em Blue curve}: The final equilibrium Fermi function at large distances. 
For small transparencies its width scales as $\sqrt{T}\Delta\mu$.}
\vspace{-3mm} \label{fig-scales}\end{center}
\end{figure}

Before reaching eventual equilibrium form, the distribution function evolves via several steps, where
its energy derivative acquires a Lorentzian shape: 
\begin{equation}
 \frac{\partial f(\epsilon)}{\partial\epsilon} = \frac{\Gamma/ \pi}{(\epsilon - \epsilon_0)^2 + \Gamma^2}.
 \quad |\epsilon|\lesssim\Delta\mu,
 \label{final-r}
\end{equation}
Here  the width $\Gamma$ and centering $\epsilon_0$ take different values in different regimes. 
Each of the regimes, summarized below and illustrated in Fig.\ \ref{fig-scales},  has its own dominant process:

(i) First, after tunneling through the QPC, electrons excite plasmons, which then split in two eigenmodes: 
one is charged fast mode with the speed $u$, and the other is slow dipole mode with the speed $v$.
This process takes place at distances $L_{\rm ex}=v/\Delta\mu$, where $\Delta\mu$ is the voltage bias 
across the QPC. In this regime Eq.\ (\ref{corr-short}) applies, which eventually leads to the the 
distribution (\ref{final-r}) with the width $\Gamma=\Gamma_{\rm ng} = 2\Delta\mu T/\pi$, centered at $\epsilon_0 =0$. 

(ii) Next, a weak dispersion of plasmons, e.g., of the form $k=\omega/v+\gamma\omega^2{\rm sign}(\omega)$, 
leads to broadening of wave-packets of the energy width $\epsilon$ and to their overlap. This process takes
place at distances $L \gg 1/\gamma \epsilon^2$. As a result, high-order cumulants of the current injected 
at the QPC are suppressed at distances $L \gg L_{\rm g}= 1/\gamma (T\Delta\mu)^2$, the noise becomes Gaussian, 
and the derivative of the electron distribution function acquires the shape (\ref{final-r}) with  the 
width $\Gamma= \Gamma_{\rm g}  = \pi \Delta\mu T/2$, centered at $\epsilon_0 = \Delta\mu T$.
 
 (iii) A situation is possible, where the dispersion of one mode, most likely of the charged mode, 
 is much stronger than the dispersion of the second mode, i.e., $\gamma_1 \gg \gamma_2$. In this case, 
 the previously described regime splits in two separate regimes. First, at distances 
 $L= 1/\gamma_1(T\Delta\mu)^2$ the contribution of the charged mode 
 to high-order cumulants of noise become suppressed, which leads to the distribution (\ref{final-r}) with 
 the parameters $\Gamma = (\Gamma_{\rm ng} + \Gamma_{\rm g})/2$ and $\epsilon_0 = \Delta\mu T/2$. Then, 
 at longer distances $L= 1/\gamma_2(T\Delta\mu)^2$ the noise becomes fully Gaussian.

 (iv) The interaction may lead to broadening of the wave-packets, but they do not decay, which implies 
 that the interaction alone does not lead to the equilibration. This means that a different, weaker 
 process may lead to the  equilibration at distances $L_{\rm eq}$ much longer than the discussed above 
 length scales. In the tunneling regime, $T\ll 1$ the width  of the eventual equilibrium distribution 
 scales as $\sqrt{T}$, in contrast to the 
 above regimes, where it scales as $T$. Thus, to observe the described variety of regimes, we propose 
 to perform measurements at large voltage biases and low transparencies of the QPC utilized to inject electrons. 
 
 Finally, we suggest a possible explanation of the paradox of missing heat flux in the experiment 
 [\onlinecite{exper2}]. So far we have summarized the effects of weak dispersion, which lead to 
 appearance of intermediate length scales. We have found that 
 in the case of a strongly nonlinear dispersion of plasmons, the measured heat flux $I_{\rm m}$
 in the outmost edge channel, 
 experimentally determined  with the procedure described by Eq.\ (\ref{flux-def}), is different from 
 the actual heat flux per channel $I_{\rm h}/2$, defined by Eq.\ (\ref{actual-f}). The screened Coulomb interaction 
 leads to a rather weak dispersion of the 
 charged plasmon, and the effect is of the opposite sign, because the spectrum in this case is concave.  
 Nevertheless, other mechanisms of the convex dispersion are possible. They will be considered elsewhere.

\begin{acknowledgments}
We would like to thank Fr\'ed\'eric Pierre for the clarification of 
the experimental details and Pascal Degiovanni for fruitful discussions.  
This work has been supported by the Swiss National Science Foundation.
\end{acknowledgments}

\appendix

\section{Solution of equations of motion}
\label{ap-nonlin}

In this appendix we solve the equations of motion (\ref{eoma}) with the boundary conditions 
(\ref{eomb}) in the case of a potential $U_{\alpha\beta}(x-y)$ of a finite range,  
where the plasmon spectrum is non-linear. 
For doing this, we first write down the normal mode expansion for the edge boson fields:
\begin{eqnarray}
\label{phi-ops}
\phi_\alpha(x) &=& \varphi_\alpha + 2\pi\cdot \pi_\alpha x\nonumber\\
&+& \sum_k \sqrt{\frac{2\pi}{kW}}[a_{\alpha k}e^{ikx} + a^\dag_{\alpha k}e^{-ikx}].
\end{eqnarray}
We consider zero modes to be classical variable, because the commutator
$[\pi_\alpha,\varphi_\alpha]=i/W$ vanishes in the thermodynamic limit $W\to 0$.
Then, we rewrite the operators $a_{\alpha k}$ in the new basis
$\tilde{a}_{j k}$, which diagonalizes the edge Hamiltonian (\ref{hamilt}):
\begin{equation}
\tilde{a}_{jk}(t) = \tilde{a}_{jk}e^{-i\omega_j(k)t},
\label{diag-ops}
\end{equation}
where $j=1,2$, and $\omega_j(k)$ is the dispersion of the $j$th mode.
In the case, where the in-channel interaction strength is approximately equal 
to the intra-channel one, $U_{\alpha\beta}(x-y)\approx U(x-y)$, and the interaction
is strong, $U(k)=\int dx e^{ikx}U(x)\gg 2\pi v_F$, the transformation to the new basis 
is simple and universal:
\begin{subequations}
\label{transf-ops}
\begin{eqnarray}
\label{a}
a_{1k}(t) = \frac{1}{\sqrt{2}}\,\big(\tilde{a}_{1k}e^{-i\omega_1(k)t}+ \tilde{a}_{2k}e^{-i\omega_2(k)t}\big),\\
\label{b}
a_{2k}(t) = \frac{1}{\sqrt{2}}\,\big(\tilde{a}_{1k}e^{-i\omega_1(k)t}- \tilde{a}_{2k}e^{-i\omega_2(k)t}\big).
\end{eqnarray}
\end{subequations}
Then, $\omega_1(k) = k[v_F+U(k)/\pi]$ for the charged plasmon,  
and $\omega_2(k) = vk$ for the dipole mode, where $v\approx v_F$.

In the next step, we use the boundary conditions (\ref{eomb})
to connect the current operators to the operators (\ref{diag-ops}):
\begin{subequations}
\label{boundary-op}
\begin{eqnarray}
\tilde{a}_{1k} = 
\frac{i}{\omega_1}\frac{\partial\omega_1}{\partial k}\sqrt{\frac{\pi k}{W}}\,[j_1(\omega_1)+j_2(\omega_1)],\\
\tilde{a}_{2k} = 
\frac{i}{\omega_2}\frac{\partial\omega_2}{\partial k}\sqrt{\frac{\pi k}{W}}\,[j_1(\omega_2)-j_2(\omega_2)],
\end{eqnarray}
\end{subequations}
where $j_\alpha(\omega) = \int dt e^{i\omega t}j_\alpha(t)$, and we have used the obvious relation
$j_\alpha(-\omega) = j^\dag_\alpha(\omega)$.
Finally, substituting relations (\ref{boundary-op}) into Eq.\ (\ref{a}) and then to the expansion 
(\ref{phi-ops}), we find the solution of the equations of motion for the boson fields. In particular,
\begin{multline}
\phi_1(x,t) = -2\pi\langle j_1\rangle t +\frac{i}{2}\int\limits_{-\infty}^\infty \frac{d\omega}{\omega}
\big\{j_1(\omega)\big(e^{ik_1x}+e^{ik_2x}\big) 
\\ + j_2(\omega)\big(e^{ik_1x}-e^{ik_2x}\big)
\big\}e^{-i\omega t},
\label{phi-disper}
\end{multline}
where we set $k_j(-\omega)=-k_j(\omega)$. In addition, we have omitted the contribution of the zero mode
$\pi_1$, because we need local correlators, and replaced the zero mode $\varphi_1(t)$ by its
expectation value.

\section{Evaluation of the measured heat flux}
\label{ap-energy-ap}

The experimentally found heat flux, defined in Eq.\ (\ref{flux-def}), may be written in 
time representation as 
\begin{equation}
I_{\rm m} =-i\partial_t
\left[K(t)-\frac{e^{i\langle\Delta\mu\rangle t}}{2\pi it}\right]_{t=0}.
\label{ener-tot}
\end{equation}
Then, we may use Eq.\ (\ref{distr-phi-2}) and results of the Appendix \ref{ap-nonlin} in order 
to evaluate the correlation function $K(t)$. The difficulty of finding this function
is related to the fact that according to Eq.\ (\ref{phi-disper}) the correlators of the operator 
$\phi_1(t)$ are determined by the currents $j_\alpha(t)$, which are
in general non-Gaussian. However, in the present case we need to take the limit $t\to 0$
in Eq.\ (\ref{ener-tot}). The high-order cumulants of currents originate from 
a non-equilibrium process and are suppressed at short times $t\Delta\mu\ll 1$.
Therefore, we are allowed to evaluate $K(t)$ in Gaussian approximation.

This may be done by expanding the right hand side of the Eq.\ (\ref{distr-phi-2})
to second order in $\phi_1$:
\begin{equation}
\ln K(t)=\langle[\phi_1(t)-\phi_1(0)]\phi_1(0)\rangle+2\pi i\langle j_1\rangle t,
\label{corr-1}
\end{equation}
where the averaging is over the fluctuations of the currents $j_\alpha$.
We then use the Eq.\ (\ref{phi-disper}) and the stationary correlators of the currents
\begin{equation}
\langle j_\alpha(\omega_1)j_\beta(\omega_2)\rangle=
2\pi S_\alpha(\omega_1)\delta_{\alpha\beta}\delta(\omega_1+\omega_2)
\label{def-stat}
\end{equation} 
to present the electron correlation function in the following form
\begin{equation}
\ln K=\ln K_{\rm n}
+2\pi i\langle j_1\rangle t-\ln t.
\label{corr-2}
\end{equation}
 Here the fluctuation contribution reads
\begin{multline}
\ln K_{\rm n}=\pi\int\limits_{-\infty}^\infty\frac{d\omega}{\omega^2}(e^{-i\omega t}-1)
\{{\mathbb S}_1(\omega)[1+\cos(\Delta kL)]\\+{\mathbb S}_2(\omega)[1-\cos(\Delta kL)]\},
\label{corr-n1}
\end{multline}
where we have introduced the excess
noise spectral densities 
${\mathbb S}_\alpha(\omega)=S_\alpha(\omega)-\omega\theta(\omega)/2\pi$
and $\Delta k =k_1(\omega)-k_2(\omega)$.

It is easy to see that for a non-vanishing contribution to $I_{\rm m}$, given by the expression
(\ref{ener-tot}), we need to expand 
$\ln K_{\rm n}$ to second order in $t$. Note, however, that linear in $t$ term in this expansion
adds to the corresponding term in Eq.\ (\ref{corr-2}) to give $2\pi i\langle\Delta\mu\rangle t$.
This follows directly from the definition (\ref{sum-rule1}) of the ``average'' bias,
which in time representation may be written as $\langle\Delta\mu\rangle=2\pi\partial_t[tK(t)]_{t=0}$.
Therefore, only $t^2$ term in $\ln K_{\rm n}$ contributes to $I_{\rm m}$, and we obtain 
\begin{equation}
I_{\rm m}=-\frac{1}{4\pi}\partial^2_t \ln K_{\rm n}(t)|_{t=0}.
\label{Im-res}
\end{equation}  
Finally, we use Eq.\ (\ref{corr-n1}) and obtain the result (\ref{tot-energy}).

Next, we wish to rewrite the measured flux $I_{\rm m}$ in terms of the 
plasmon distributions $n_j(k)=\langle \tilde a^\dagger_j(k)\tilde a_j(k)\rangle$, $j=1,2$. 
For doing so, we now use Eqs.\ (\ref{phi-ops}) and (\ref{transf-ops}), repeat the steps that 
lead to (\ref{corr-n1}), and take the limit of $L\gg L_{\rm ex}$. The result
may be presented in the following form: 
\begin{eqnarray}
\ln K_{\rm n}=&-&\sum_j\int\limits_0^\infty \frac{dk}{k}\,n_j(k)[1-\cos(\omega_jt)]\nonumber\\
&+&\frac{1}{2}\sum_j\int\limits_0^\infty \frac{dk}{k}\,\left(e^{-i\omega_jt}-e^{-iv_jkt}\right),
\label{corr-n2}
\end{eqnarray}
where the last term is the quantum contribution due to the non-linear plasmon spectrum,
and $v_j=\partial\omega_j/\partial k$ are the plasmon speeds at $k=0$. Substituting expressions (\ref{corr-n2}) 
into Eq.\ (\ref{Im-res}), we obtain the final result (\ref{measured-f}) 
and (\ref{quantum-f}) for the measured heat flux.

\section{Suppression of higher order cumulants}
\label{ap-supr}

Here we show that a weak dispersion of plasmon modes leads to the suppression of the contribution of higher order 
cumulants of current $j_\alpha$ to the electron correlation function at long distances $L$ from the source of 
currents. We demonstrate this using an example of a weakly dispersive spectrum of plasmons in the form
$k_j = \omega/v_j +\gamma_j\omega^2{\rm sign}(\omega)$, $j=1,2$. Since we are interested in the behavior of the 
electron distribution function close to the Fermi level, we need to know a long-time asymptotics of the 
electron correlation function $K(t)$. Therefore, the contributing currents $j_\alpha$ can be considered 
Markovian processes and fields 
$\phi_\alpha$ can be treated as classical variables.

Let us consider the $n$th cumulant: 
\begin{equation}
M^{(n)}(L,t)\equiv \langle\!\langle[\phi_1(L,t)-\phi_1(L,0)]^n\rangle\!\rangle.
\end{equation} 
According to Eq.\ (\ref{phi-disper}), at large distances $L\gg L_{\rm ex}=v_j/\Delta\mu$ 
and long times  $t\Delta\mu \gg 1$ 
it may be written as
\begin{equation} 
M^{(n)}(L,t) = \sum_\alpha M^{(n)}_\alpha(L,t),
\label{M-n}
\end{equation} 
where:
\begin{multline}
\label{xxx}
M^{(n)}_\alpha(L,t) = 2\pi S_\alpha^{(n)}\int \prod_{l=1}^n\frac{ d\omega_l}{\omega_l}(i/2)(e^{-i\omega_l t}-1)
\\ \times\delta(\omega_1+\ldots+\omega_n)\Big[e^{i\sum_lk_1 (\omega_l)L}+e^{i\sum_lk_2 (\omega_l)L}\Big]
\end{multline}
and $S_\alpha^{(n)}\equiv \langle\!\langle j_\alpha^n\rangle\!\rangle$. Here  we have neglected the cross 
terms containing fast oscillating functions. These terms
have the same origin as fast oscillating terms in (\ref{tot-energy}) and 
vanish at distances $L\gg L_{\rm ex}$. Dropping those terms is also equivalent 
to neglecting in (\ref{corr-long}) correlations of charges taken at different times $t_u$ and $t_v$.
Finally, we note that in our particular case, where the QPC is connected to the outmost edge channel
only, $S_2^{(n)}=0$ for $n>2$. 

One can easily see that $\sum_l k_j(\omega_l) = \sum_l \gamma_j\omega_l^2{\rm sign}(\omega_l)$, 
because the integrals in (\ref{xxx}) are limited to $\sum_l \omega_l=0$. For the second cumulant
this implies that $k_j(\omega_1)+k_j(\omega_2)=0$, i.e., the dispersion correction cancels too. 
Therefore the second cumulant is not suppressed at long distances. Below we consider high-order cumulants.
Using the identity $2\pi\delta(\omega_1+\ldots+\omega_n) = \int d\tau \exp [i(\omega_1+\ldots+\omega_n)\tau]$ 
we can write
\begin{equation}
 M^{(n)}_\alpha(L,t) = S_\alpha^{(n)}\sum_j\!\int\limits_{-\infty}^\infty d\tau [F_j(\tau, t, L)]^n, 
 \label{integral-F}
\end{equation}
where we have introduced the integrals
\begin{equation}
F_j = \frac{i}{2}\int \frac{d\omega}{\omega} (e^{-i\omega t}-1)e^{i\omega \tau + i\gamma_jL \omega^2{\rm sign}(\omega) }.
\label{ha-ha}
\end{equation}
At large distances $L\gamma_j \gg t^2$ the  contribution to the integrals $F_j$ comes from small $\omega$, 
where one can approximate $e^{-i\omega t}-1\approx -i\omega t$. 
Therefore, Eq.\ (\ref{ha-ha}) can be further simplified:
\begin{equation}
 F_j = (t/2)\int d\omega e^{i\omega \tau + i\gamma_jL \omega^2{\rm sign}(\omega)}  \propto \frac{t}{\sqrt{\gamma_j L}}e^{\pm i\tau^2/4\gamma_j L}. 
\end{equation}
Substituting this result into Eq.\ (\ref{integral-F}) and then to (\ref{M-n}), we find that
\begin{equation}
M^{(n)}(L,t)\propto \,t\sum_\alpha S_\alpha^{(n)}\!\sum_{j=1,2} \Big(\frac{t^2}{\gamma_jL}\Big)^{\frac{n-1}{2}},\; n>2,
\label{result-M}
\end{equation}
where, we recall, the sum is over the plasmon eigenmode number $j$ and over the channel number $\alpha$. 

We note, that at large distances $L$ the cumulants $M^{(n)}(L,t)$ are suppressed by 
the dimensionless small parameter $t^2/\gamma_jL\ll 1$. In our case 
$S_1^{(n)}\sim T\Delta\mu$ and $S_2^{(n)}=0$, therefore the contribution of high-order cumulants 
to the correlator $K(t)$ may be neglected 
at distances larger than $L_{\rm g}=1/\gamma_j(T\Delta\mu)^2$, and noise may be considered 
Gaussian. Obviously, if only one plasmon 
mode of two is dispersive, e.g., $\gamma_2=0$, then at distances $L\gg L_{\rm g}$ 
the cumulant (\ref{M-n}) 
is suppressed by the factor of 2. One can interpret the result (\ref{result-M}) as 
the renormalization of the effective coupling constant $\lambda$ in (\ref{fcs}), which 
is caused by spreading of a plasmon wave packets due to the dispersion. 

\bibliographystyle{apsrev}

\begin{thebibliography}{99}

\bibitem{exper}
C. Altimiras {\em et al.}, Nature Physics {\bf 6}, 34 (2010).

\bibitem{exper2}
H. le Sueur {\em et al.}, Phys. Rev. Lett. {\bf 105}, 056803 (2010);
C. Altimiras {\em et al.}, Phys. Rev. Lett. {\bf 105}, 226804 (2010).

\bibitem{nigg}
A.M. Lunde, S.E. Nigg, M. Buttiker, Phys. Rev. B {\bf 81}, 041311(R) (2010).

\bibitem{giovan}
P. Degiovanni {\em et al.}, Phys. Rev. B {\bf 81}, 121302(R)(2010).

\bibitem{our-phas}
I.P. Levkivskyi, E.V. Sukhorukov, Phys. Rev. Lett. {\bf 103}, 036801 (2009).

\bibitem{qhe}
K. v. Klitzing, G. Dorda, and M. Pepper, Phys. Rev. Lett. {\bf 45}, 494 (1980).

\bibitem{firstMZ}
Y.\ Ji {\em et al}., Nature (London) {\bf 422}, 415 (2003).

\bibitem{Heiblum}
I. Neder {\em et al}., Phys. Rev. Lett. {\bf 96}, 016804 (2006);
I. Neder {\em et al}., Nature Physics {\bf 3}, 534 (2007).

\bibitem{Glattli}
P. Roulleau {\em et al.}, Phys. Rev. B {\bf 76}, 161309(R) (2007).
P. Roulleau {\em et al.}, Phys. Rev. Lett. {\bf 100}, 126802 (2008).

\bibitem{Litvin}
L.V. Litvin {\em et al.}, Phys. Rev. B {\bf 75}, 033315 (2007).
L.V. Litvin {\em et al.}, Phys. Rev. B {\bf 78}, 075303 (2008).


\bibitem{Basel}
E. Bieri {\em et al.}, Phys. Rev. B {\bf 79}, 245324 (2009).

\bibitem{group1}
B.I. Halperin, Phys. Rev. B {\bf 25}, 2185 (1982); M. B\"uttiker, 
Phys. Rev. B {\bf 38}, 9375 (1988) and references therein.

\bibitem{group2}
V.A. Volkov and S.A. Mikhailov, Sov. Phys. JETP {\bf 67}, 1639 (1988); D.B.
Chklovskii, B.I. Shklovskii, and L.I. Glazman, Phys. Rev. B {\bf 46}, 4026 (1992); 
I.L. Aleiner and L.I. Glazman, Phys. Rev. Lett. {\bf 72}, 2935 (1994); 
C.d.C. Chamon and X.G. Wen, Phys. Rev. B {\bf 49}, 8227 (1994) and references therein.

\bibitem{eff-theory}
X.-G. Wen, Phys. Rev. B {\bf 41}, 12838 (1990);
J. Fr\"{o}hlich, A. Zee, Nucl. Phys. B{\bf 364}, 517 (1991).

\bibitem{Chang}
For a review, see
A.M. Chang, Rev. Mod. Phys. {\bf 75}, 1449 (2003).

\bibitem{Sukh-Che}
E.V. Sukhorukov, V.V. Cheianov, Phys. Rev. Lett. {\bf 99}, 156801 (2007).

\bibitem{Chalker}
J.T. Chalker, Y. Gefen, and M.Y. Veillette,
Phys. Rev. B {\bf 76}, 085320 (2007).

\bibitem{our}
I.P. Levkivskyi, E.V. Sukhorukov, Phys. Rev. B {\bf 78}, 045322 (2008).

\bibitem{Neder}
I. Neder, E. Ginossar, Phys. Rev. Lett. 100, 196806 (2008).

\bibitem{Sim}
S.-C. Youn, H.-W. Lee, and H.-S. Sim, Phys. Rev. Lett. {\bf 100}, 196807 (2008).

\bibitem{1d}
C.L. Kane, M.P.A. Fisher, Phys. Rev. B {\bf 52}, 17393 (1995); 
V.V. Ponomarenko, D.V. Averin , Phys. Rev. B {\bf 67}, 035314 (2003); 
D. B. Gutman, Y. Gefen, A. D. Mirlin, Eur. Phys. Lett. {\bf 90}, 37003 (2010); 
D. L. Kovrizhin, J. T. Chalker, Phys. Rev. B {\bf 84}, 085105 (2011).

\bibitem{1d-exp}
L.P. Kouwenhoven, B.J. van Wees, N.C. der Vaart, C.J.P.M. Harmans, C.E.
Timmering, C.T. Foxon, Phys. Rev. Lett. {\bf 64}, 685 (1990); 
Pothier H., Guéron S., Birge N.O., Esteve D., Devoret M.H., Phys. Rev. Lett. {\bf 79},
3490 (1997);
Anthore A., Pierre F., Pothier H., Esteve D., Phys. Rev. Lett. {\bf 90}, 076806
(2003);
Granger G., Eisenstein J.P.,
Reno J.L., Phys. Rev. Lett. {\bf 102}, 086803 (2009).

\bibitem{citeus}
In fact, exactly same physical effect leads to the decoherence in MZ interferometers \cite{Heiblum,Basel,
Glattli,Litvin}
and is responsible for the anomalous dephasing, as demonstrated in our earlier work [\onlinecite{our-phas}].

\bibitem{Levitov} L.S. Levitov, H. Lee, and G.B. Lesovik, J.\ Math.\
Phys. {\bf 37}, 4845 (1996).

\bibitem{Giamarchi}
Th. Giamarchi, {\em Quantum Physics in One Dimension}
(Oxford University Press, Oxford, 2003).

\bibitem{Ines} Note, that the problem of finding the conductance of a 1D
system attached to ohmic reservoirs requires a different type of 
boundary conditions for fields $\phi_\alpha$, imposed at reservoirs, where 
currents are classical variables. See, e.g.,   
I. Safi, Eur. Phys. J. B. {\bf 12}, 451 (1999).

\bibitem{dynamicalCB} For a recent experiment, see
C. Altimiras {\em et al.},
Phys. Rev. Lett. {\bf 99}, 256805 (2007).

\bibitem{nagaev}
K. E. Nagaev, Phys. Rev. B {\bf 66}, 075334 (2002).

\bibitem{footnote1} We would like to reiterate the role of the interaction at 
filling factor $\nu=2$ by comparing to the situation at $\nu =1$. In the last case, 
there is only one channel at the edge, so that the solution of the equation of 
motion acquires the form $\phi(x, t) = 2\pi Q(t-x/u)$, and the correlation 
function (\ref{corr-long}) is given by $\chi(2\pi,t)$. One can show that this 
correlation function describes free electrons, therefore the screened Coulomb 
interaction leads solely to the renormalization of the the Fermi velocity.

\bibitem{Edwards}
E. V. Sukhorukov, and J. Edwards, Phys. Rev. B {\bf 78}, 035332 (2008).

\bibitem{FDT}
H.B. Callen, and T.A. Welton, Phys. Rev. {\bf 83}, 34 (1951).

\bibitem{Buttiker}
For a review, see Y.M. Blanter, M. B\"uttiker, Phys. Rep. {\bf 336}, 1 (1986).

\bibitem{footnote2}
We choose this example of the dispersion correction
in the spectrum merely to do estimates, and to draw general conclusions, 
and not as a realistic example. 

\bibitem{footnote3}
We note, that in general the heat flux cannot be associated with a conserved Noether 
current, because the Hamiltonian (\ref{Hamiltonian}) is not local. Nevertheless, the 
continuity equation for the heat flux $I_{\rm h}$, given by (\ref{actual-f}), may be
derived in the semi-classical limit $kv<\Delta\mu$, which is of interest here.
Moreover, one can show that for any interaction matrix (\ref{kernel}) 
the energy flux of zero modes of plasmons does not contribute to the total heat flux,
if it is defined as the Joule heat created at the QPC.




\end{thebibliography}

\end{document}